\documentclass[a4paper,11pt]{article}
\pdfoutput=1 

\usepackage{jheppub} 

\usepackage[T1]{fontenc} 

\usepackage[colorlinks=true]{hyperref}

\usepackage{siunitx}
\usepackage{graphicx}
\usepackage{amsmath}
\usepackage{float}
\usepackage{caption}
\usepackage{subcaption}
\usepackage{multirow}
\usepackage{xspace}
\usepackage[switch]{lineno}

\newcommand{\ampgen}{\texttt{AmpGen}}
\newcommand{\kspipi}{\ensuremath{K_S^0 \pi^+ \pi^-}\xspace}
\newcommand{\Dz}{\ensuremath{D^0}\xspace}
\newcommand{\Dzbar}{\ensuremath{\overline{D}{}^0}\xspace}
\newcommand{\ks}{\ensuremath{K_S^0}\xspace}
\newcommand{\KS}{\ks}
\newcommand{\psipp}{\ensuremath{\psi(3770)}\xspace}
\newcommand{\phid}{\ensuremath{\phi_D}\xspace}
\newcommand{\phidbar}{\ensuremath{\overline{\phi}_D}\xspace}
\newcommand{\dd}{\ensuremath{\delta_D}\xspace}
\newcommand{\ddbar}{\ensuremath{\overline{\delta}_D}\xspace}
\newcommand{\ddmodel}{\ensuremath{\dd^{\mathrm{model}}}\xspace}
\newcommand{\ddcorr}{\ensuremath{\dd^{\mathrm{corr}}}\xspace}
\newcommand{\mplus}{\ensuremath{s_+}\xspace}

\newcommand{\mminus}{\ensuremath{s_-}\xspace}

\newcommand{\myCP}{\ensuremath{\mathit{CP}}\xspace}
\newcommand{\myC}{\ensuremath{\mathit{C}}\xspace}

\newcommand{\ket}[1]{\ensuremath{|#1\rangle}}

\newcommand{\psp}{\ensuremath{\mathbf{p}}\xspace}
\newcommand{\pspbar}{\ensuremath{\mathbf{\overline{p}}}\xspace}
\newcommand{\psq}{\ensuremath{\mathbf{q}}\xspace}

\newcommand{\fD}{\ensuremath{f}}
\newcommand{\prt}[1]{\ensuremath{#1}}
\newcommand{\ADz}{\ensuremath{A_{D}}\xspace}
\newcommand{\ADzbar}{\ensuremath{\overline{A}_{D}}\xspace}
\newcommand{\ADzf}{\ensuremath{A_{D}^{\fD}}\xspace}
\newcommand{\ADzbarf}{\ensuremath{\overline{A}_{D}^{\fD}}\xspace}
\newcommand{\Real}{\ensuremath{\mathsf{Re}}}

\newcommand{\eqnref}[1]{Eq~\ref{#1}}

\newcommand{\figref}[1]{Fig~\ref{#1}}
\newcommand{\Tabref}[1]{Table~\ref{#1}}
\newcommand{\tabref}[1]{Tab~\ref{#1}}
\newcommand{\Secref}[1]{Section~\ref{#1}}
\newcommand{\secref}[1]{Sec~\ref{#1}}

\newcommand{\besIII}{BES~III\xspace}
\newcommand{\cleoc}{\mbox{CLEO-c}\xspace}
\newcommand{\babar}{BaBar\xspace}
\newcommand{\belle}{Belle\xspace}
\newcommand{\belleII}{Belle~II\xspace}
\newcommand{\hide}[1]{}
\newcommand{\te}{\ensuremath{\mbox{}^{\mathrm{th}}}\xspace}

\title{\boldmath A novel unbinned model-independent method to measure the CKM angle $\gamma$ in $B^\pm \to DK^\pm$ decays with optimised precision.}

\author[a,b]{Jake Lane,}
\author[b]{Evelina Gersabeck,}
\author[c]{and Jonas Rademacker}

\affiliation[a]{Monash University,\\Melbourne, Australia}
\affiliation[b]{University of Manchester,\\Manchester, United Kingdom}
\affiliation[c]{University of Bristol,\\Bristol, United Kingdom}

\emailAdd{jake.lane@cern.ch}
\emailAdd{evelina.gersabeck@cern.ch}
\emailAdd{jonas.rademacker@bristol.ac.uk}

\abstract{
We present a novel unbinned method to combine \prt{B^{\pm} \to DK^{\pm}} and charm threshold data for the amplitude-model unbiased measurement of the CKM angle $\gamma$ in cases where the \prt{D} meson decays to a three-body final state. The new unbinned approach avoids any kind of integration over the \prt{D} Dalitz plot, to make optimal use the available information. We verify the method with simulated signal data where the \prt{D} decays to \prt{K_S \pi^+ \pi^-}. Using realistic sample sizes, we find that the new method reaches the statistical precision on $\gamma$ of an unbinned model-dependent fit, i.e. as good as possible and better than the widely used model-independent binned approach, without suffering from biases induced by a mis-modeled \prt{D} decay amplitude.
\vfill{}\noindent
\centerline{
Published in \href{https://doi.org/10.1007/JHEP09(2023)007}{JHEP 09 (2023) 007}.}
}

\begin{document} 

\maketitle

\flushbottom

\section{Introduction}
The precision measurement of the \myCP violating phase $\gamma$ ($\phi_3$) is one of the primary goals of flavour physics today.
The measurement of $\gamma$ in \prt{B^{\pm} \to DK^{\pm}} and related decays~\cite{Gronau:1991dp,Gronau:1990ra,Bondar:2002unpublished, GGSZ, Belle:2004bbr,Atwood:1996ci} has negligible theoretical uncertainty~\cite{Brod:2013sga}. The current precision on $\gamma$ is dominated by LHCb, which measures $\gamma = \left(63.8^{+3.5}_{-3.7}\right)^{\circ}$~\cite{LHCb:2022awq, LHCb:2021dcr}. The vast, clean datasets expected from the recently commissioned LHCb upgrade~\cite{Bediaga:1443882} and \belleII~\cite{Belle-II:2010dht} will allow a sub $1^{\circ}$ precision on $\gamma$, LHCb upgrade~II is expected to reduce this further to $0.35^{\circ}$~\cite{LHCbCollaboration:2776420, LHCb:2018roe}.
The parameter $\gamma$ is therefore set to become the most precisely measured parameter in the Cabibbo-Kobayashi-Maskawa (CKM) description of \myCP violation in the quark sector~\cite{CofCKM, KMofCKM}, giving it a pivotal role in the search for new physics by over-constraining the Standard Model with precision measurements. To fully benefit from this potential, it is critical to control systematic uncertainties.

The measurement of $\gamma$ in \prt{B^{-} \to DK^{-}} and its \myCP conjugate is possible because the neutral \prt{D} meson in this decay is a superposition of \Dz and \Dzbar that depends on $\gamma$: \prt{D \propto \Dz + r_B e^{i(\delta_B - \gamma)} \Dzbar}, where $r_B\sim 0.1$ and $\delta_B$ is a \myCP-conserving phase induced by the strong interaction. The neutral \prt{D} meson is reconstructed in a final state accessible to both \Dz and \Dzbar. 
For multibody decays of the \prt{D}, such as \prt{\KS\pi^+\pi^-}, the \myCP-violating phase $\gamma$ is obtained from analysing the amplitude structure of the \prt{D} decay in both \prt{B^{-} \to DK^{-}} and \prt{B^{+} \to DK^{+}} transitions; this approach is known as the BPGGSZ method~\cite{Bondar:2002unpublished, GGSZ, Belle:2004bbr}.

In order to measure $\gamma$ in this way, 
the phase difference between the \Dz and \Dzbar decay amplitude across the Dalitz plot needs to be known.
While this information can be obtained from amplitude models, these models have well-known shortcomings that make the phase information they provide unreliable, leading to significant, difficult-to-quantify systematic uncertainties. For this reason, model-independent methods are required to achieve the ultimate precision on $\gamma$. Similar considerations apply to measurements of \prt{D}-mixing~\cite{Bondar:2010qs}.


Model-independent methods currently in use rely on integrating over all or parts (bins) of the multibody phase space of the \prt{D} decay~\cite{GGSZ, Atwood:2003mj, Bondar:2005ki, Harnew:2013wea, Harnew:2014zla, Nayak:2014tea}. Two unbinned model-independent methods have been proposed recently. The method described in~\cite{Poluektov:2017zxp} is based on projecting the two-dimensional Dalitz plot down to one dimension, where the phase difference between the \Dz and \Dzbar amplitudes is parameterised as a Fourier series. The authors of~\cite{Backus:2022xhi} extract $\gamma$ from set of cumulative functions defined across the Dalitz plot in a way that is inspired by the Kolmogorov-Smirnov test. This method is independent of the phase difference between the \Dz and \Dzbar amplitudes.

In all cases, one can expect some information loss due to the integration or projection process involved. Here we present a new unbinned model-independent method that optimally uses the full information across the two-dimensional Dalitz plot of a three-body decay of the neutral \prt{D} meson. Using the example of \prt{\Dz \to \KS \pi^+ \pi^-}, we will show that this method has, with current and plausible future data sample sizes, the potential to reach essentially the same statistical precision on $\gamma$ as a model-dependent method, without suffering from the associated model uncertainty.

This paper is organised as follows: In \secref{sec:formalism} we remind the reader of the formalism for the measurement of $\gamma$ in \prt{B^{\pm} \to DK^{\pm}} decays, and use this opportunity to introduce our notation and phase convention. \Secref{sec:qmiMethodTheory} describes the new quasi model-independent method introduced in this paper. In \secref{sec:simulationStudies}, we evaluate the performance of the new method in simulation studies for the measurement of $\gamma$ in \prt{B^{\pm} \to DK^{\pm}, D \to \KS \pi^+\pi^-} decays. Finally, in \secref{sec:conclusion}, we conclude.
\section{Formalism}
\label{sec:formalism}
In this section we outline the formalism for the measurement of $\gamma$ and the use of charm threshold data developed in~\cite{Gronau:1991dp,Gronau:1990ra,Bondar:2002unpublished, GGSZ, Belle:2004bbr,Atwood:1996ci}, to remind the reader and fix the notation. For simplicity, we ignore the effects of \prt{D} mixing~\cite{Meca:1998ee, Silva:1999bd, Grossman:2005rp, Rama:2015pmr}. 
In most cases, these effects are small enough to be negligible~\cite{Grossman:2005rp}. Where they are not, they can be taken into account as described in~\cite{Rama:2015pmr}.

\subsection{Notation and conventions}
\label{sec:notation}
We use the notation $\ADzf(\psp)$ for the decay amplitude of the \Dz meson to a final state \fD, at point $\psp$ in the \prt{D \to \fD} Dalitz plot and $\ADzbarf(\pspbar)$ for that of the \myCP-conjugate process. 

Our phase convention for the \myCP\ operator is such that $\myCP\ket{\Dz} = \ket{\Dzbar}$, which is the usual practice in the context of beauty decays to charm. An alternative convention where $\myCP\ket{\Dz} = - \ket{\Dzbar}$ is widely used in the context of charm physics.
 
With our phase convention, and assuming the absence of \myCP violation in charm decays, $\ADzf(\psp)=\ADzbarf(\pspbar)$.
For self-conjugate final states such as \prt{\KS \pi^+ \pi^-}, which we will use for the simulation studies presented below, $\psp$ and $\pspbar$ are in the same Dalitz plot. We parameterise the \prt{\Dz \to \KS\pi^+\pi^-} Dalitz plot with the usual variables $s_+ = m^2(\KS \pi^+)$ and $s_- = m^2(\KS \pi^-)$, representing the invariant mass-squared of the \prt{\KS\pi^+} and \prt{\KS\pi^-} pair, respectively. If $\psp = (s_-, s_+)$, then $\pspbar= (s_+, s_-)$, and $\ADzbar^{\KS\pi\pi}(s_+, s_-) = \ADz^{\KS\pi\pi}(s_-, s_+)$.

We define the phases $\phid^{\fD}(\psp) \equiv \arg(\ADzf(\psp)), \phidbar^{\fD}(\psp) \equiv \arg(\ADzbarf(\psp))$ and the phase difference $\dd^{\fD}(\psp) \equiv \phid^{\fD}(\psp) - \phidbar^{\fD}(\psp)$. 
It is also useful to define $\ddbar^{\fD}(\psp) \equiv \phidbar^{\fD}(\psp) - \phid^{\fD}(\psp)$, even though it is trivially related to \dd through $\ddbar^{\fD}(\psp) = \dd^{\fD}(\pspbar) = -\dd^{\fD}(\psp)$. To declutter the notation we will omit the superscripts and/or the $(\psp)$, where there is no risk of ambiguity.

When applying the method presented below in practice it will frequently be necessary to combine results obtained using different conventions for the phase of the \myCP operator. Switching from our convention with $\myCP\ket{\Dz} = \ket{\Dzbar}$ to the convention with $\myCP\ket{\Dz} = -\ket{\Dzbar}$ corresponds to the change $\dd \to \dd + \pi$.

\subsection[Measuring gamma in B --> DK]{Measuring $\gamma$ with \prt{B^{\pm}\to DK^{\pm}} decays}
\label{sec:gammaFormalism}
The decay amplitude of a \prt{D} meson resulting from a \prt{B^- \to DK} to a final state $\fD$ at phase space point \psp is given by
\begin{align}
A_{B-}(\psp) \propto \ADz(\psp) + r_B e^{i(\delta_B - \gamma)} \ADzbar(\psp),
\end{align}
and the corresponding decay rate is
\begin{align}
\Gamma^-(\psp) & = N \left( |\ADz(\psp)|^2 + r_B^2 |\ADzbar(\psp)|^2 + 2\Real(\ADz(\psp) r_Be^{-i(\delta_B(\psp) - \gamma)}\ADzbar^*(\psp)) \right)
  \\    & = N\left(|\ADz(\psp)|^2 + r_B^2 |\ADzbar(\psp)|^2 + 2r_B |\ADz(\psp)| |\ADzbar(\psp)| \cos(-\delta_B + \gamma + \dd)\right)
  \\    
  \label{eq:rateBm}
  & = N\big(|\ADz(\psp)|^2 + r_B^2 |\ADzbar(\psp)|^2 
  \nonumber \\
  &\;\;\; + 2r_B |\ADz(\psp)| |\ADzbar(\psp)| 
  \left( \cos(\delta_B - \gamma)\cos(\dd(\psp)) + \sin(\delta_B - \gamma) \sin(\dd(\psp))\right)\big),
\end{align}
where $N$ is a normalisation factor. For the \myCP-conjugate process, with a \prt{D} from a \prt{B^+ \to DK^+}:
\begin{align}
\label{eq:rateBp}
\Gamma^+(\pspbar) & = N \big( |\ADzbar(\pspbar)|^2 + r_B^2 |\ADz(\pspbar)|^2 
\nonumber\\
&\;\;\;+ 2r_B |\ADz(\pspbar)| |\ADzbar(\pspbar)| \left( \cos(\delta_B + \gamma)\cos(\ddbar) + \sin(\delta_B + \gamma) \sin(\ddbar(\pspbar)) \right) \big)
\end{align}
where $\ADzbar(\pspbar)=\ADz(\psp)$, $  \ADz(\pspbar) = \ADzbar(\psp)$, and $\ddbar(\pspbar) = \dd(\psp)$ for \myCP conservation in charm.
For the fits in our validation studies, we follow the widely used practice to re-parameterise the decay rates in terms of the ``cartesian'' variables
\begin{align}
    x_{+} &\equiv r_B  \cos(\delta_B + \gamma), &
    y_{+} &\equiv r_B  \sin(\delta_B + \gamma), &
    x_{-} &\equiv r_B  \cos(\delta_B - \gamma), &
    y_{-} &\equiv r_B  \sin(\delta_B - \gamma).
\end{align}
This is motivated by the observation that fits in terms of $x_{\pm}$ and $ y_{\pm}$ are statistically better behaved than those in terms of $r_B, \delta_B,$ and $\gamma$, which is related to the fact that there is no sensitivity to $\delta_B, \gamma$ as $r_B \to 0$. The decay rates in terms of the new variables are
\begin{align}
  \label{eq:rateBmx}
\Gamma^-(\psp)
   & = N\left(|\ADz|^2 + \left(x_-^2 + y_-^2\right) |\ADzbar|^2 + 2 |\ADz| |\ADzbar| \left(x_- \cos(\dd) + y_- \sin(\dd)\right)\right),
   \\
     \label{eq:rateBpx}
 \Gamma^+(\pspbar) & = N \left( |\ADzbar|^2 + \left(x_+^2 + y_+^2\right) |\ADz|^2 + 2|\ADz| |\ADzbar| \left( x_+ \cos(\ddbar) + y_+ \sin(\ddbar) \right) \right).
\end{align}

In a model-dependent approach,
\ADz and \ADzbar are derived from a high-statistics amplitude fit to flavour-specific \Dz, \Dzbar decays. However, the models used to describe the decay amplitude have theoretical shortcomings that make their phase information, which enters via \dd, unreliable. This in turn translates into a systematic uncertainty on $\gamma$.

For the model-independent approach described in \cite{GGSZ, Bondar:2005ki}, one integrates over regions (bins) of phase space. These bins are defined such that they form \myCP-conjugate pairs and we label them such that the \myCP-conjugate of bin $i$ is bin $-i$. In what follows, we assume the \prt{D} decays to \prt{\KS\pi^+\pi^-}, although the approach clearly generalises to other decay modes. We define the following parameters related to $|\ADz|^2, |\ADzbar|^2$:
\begin{align}
F_i &\equiv \int_{\mathrm{bin}\;i} |\ADz|^2 ds_+ ds_- 
, &
\overline{F}_i &\equiv \int_{\mathrm{bin}\;i} |\ADzbar|^2 ds_+ ds_- .
\end{align}
In the absence of \myCP\ violation in charm, $\overline{F}_i = F_{-i}$. We also define the following parameters related to the phase-difference between \ADz and \ADzbar:
\begin{align}
c_i &\equiv \frac{\int_{\mathrm{bin}\;i} |\ADz| |\ADzbar| \cos(\dd) ds_+ ds_-}{
\sqrt{\int_{\mathrm{bin}\;i} |\ADz|^2 ds_+ ds_-  \int_{\mathrm{bin}\;i} |\ADzbar|^2 ds_+ ds_-}
},
\\
s_i &\equiv \frac{\int_{\mathrm{bin\;i}} |\ADz| |\ADzbar| \sin(\dd) ds_+ ds_-}{
\sqrt{\int_{\mathrm{bin\;i}} |\ADz|^2 ds_+ ds_-  \int_{\mathrm{bin\;i}} |\ADzbar|^2 ds_+ ds_-}
},
\end{align}
which implies $c_{-i} \equiv c_i$ and $s_{-i} \equiv - s_i$.
In terms if these quantities, the decay rate \prt{B^- \to DK^-} with the \prt{D} decay in phase-space bin $i$, is given by
\begin{align}
\Gamma^-_i & =
F_i + r_B^2 {F}_{-i} + 2r_B \sqrt{F_i F_{-i}} 
\left( \cos(\delta_B - \gamma)c_i + \sin(\delta_B - \gamma) s_i
\right).
\end{align}
 The \myCP\ conjugate process, a \prt{B^+ \to DK^+} decay with the \prt{D} decay in phase-space bin $-i$, is
\begin{align}
\Gamma^+_{-i} & =  
{F}_{i} + r_B^2 {F}_{-i} + 2r_B \sqrt{F_i F_{-i}} 
\left(
\cos(\delta_B + \gamma)c_i + \sin(\delta_B + \gamma) s_i
\right).
\end{align}
Equivalently, these decay rates can be expressed in terms of $x_{\pm}, y_{\pm}$:
\begin{align}
\Gamma^-_i & =
F_i + \left(x_-^2 + y_-^2\right) {F}_{-i} + 2 \sqrt{F_i F_{-i}} 
\left( x_- c_i + y_- s_i
\right),
\\
\Gamma^+_{-i} & = 
{F}_{i} + \left(x_+^2 + y_+^2\right) {F}_{-i} + 2 \sqrt{F_i F_{-i}} 
\left(
x_+ c_i + y_+ s_i
\right).
\end{align}
All parameters related to the charm decay, i.e. $F_i, c_i$, and $s_i$, can be directly obtained from data, where data from the charm threshold are critical to constraining the parameters related to the phase difference between \ADz and \ADzbar, i.e. $c_i$ and $s_i$. 
It is worth noting, though, that the $c_i$ and $s_i$ can be obtained alongside $\gamma$ from (a sufficiently large sample of) \prt{B^{\pm} \to DK^{\pm}} decays~\cite{GGSZ}; however, the input from the charm threshold dramatically improves the fit. Measurements of $\gamma$ in this way, using the \prt{D\to \KS \pi^+ \pi^-} Dalitz plot we focus on here, have been made by \belle~\cite{Belle:2012ftx, Belle:2015roy}, \belle \&\ \belleII~\cite{Belle:2021efh}, and LHCb~\cite{LHCb:2012apu,LHCb:2018wag, LHCb:2020yot}, using input from \cleoc~\cite{CLEO:2010iul, CLEO:2009syb} and \besIII~\cite{BESIII:2020hlg, BESIII:2020hpo}.
We will below introduce a new method, also based on exploiting threshold data, 
that does not require binning, or other forms of integration over phase space as in~\cite{Poluektov:2017zxp, Backus:2022xhi}. This is motivated by the aim to maximise the use of information contained in the full two-dimensional Dalitz plot.

\subsection{Charm threshold data}
\label{sec:thresholdFormalism}
At the charm threshold, $\psi(3770)$ are produced and decay approximately 50\%\ of the time to a pair of neutral \prt{D} mesons that we label \prt{D_1, D_2}. The resulting system of \prt{D} mesons must be $\myC$-odd, like the \prt{\psi(3770)} they originate from. Therefore:
\begin{align}
\ket{\psi(3770)} \to \ket{\Dz}_1\ket{\Dzbar}_2 - \ket{\Dzbar}_1 \ket{\Dz}_2.
\end{align}
Let \prt{D_1} decay to final state $f$ at phase-space point \psp and \prt{D_2} to final state $g$ at phase-space point \psq. Then the decay amplitude for this process is
\begin{align}
A(\psi \to D_1 D_2; D_1 \to f(\psp), D_2 \to g(\psq)) &= \frac{1}{\sqrt{2}} \left(
\ADz^f(\psp)\ADzbar^g(\mathbf{q})
-
\ADzbar^f(\psp)\ADz^g(\mathbf{q})
\right).
\end{align}
We will call $g$ the \emph{tag} and $f$ the \emph{signal} (in our feasibility study, $f$ will be \prt{\KS\pi^+\pi^-}). Depending on the tag, we can distinguish a few important special cases:
\begin{enumerate}
\item \prt{D_2 \to g} is a flavour-specific decay such as a semileptonic decay, or a quasi-flavour-specific decay such as \prt{\Dz \to K^- \pi^+} (in our feasibility studies, we will ignore the small dilution effect due to the suppressed decay \prt{\Dzbar \to K^- \pi^+}, although this can be taken into account). Then
\begin{align}
A(\psi \to D_1 \Dz; D_1 \to f(\psp), \Dz \to g) & \propto \ADzbar^f(\psp),
\end{align}
and similarly
\begin{align}
A(\psi \to D_1 \Dzbar; D_1 \to f(\psp), \Dzbar \to g) & \propto \ADz^f(\psp).
\end{align}
We refer to these decays as flavour tagged. They provide the same information as flavour-tagged \Dz\ decays at the B-factories and LHCb, where the flavour is usually identified through the charge of the pion in \prt{D^{*+} \to \Dz \pi^+} and \prt{D^{*-} \to \Dzbar \pi^-} decays. Measurements of flavour-tagged decays provide $|\ADz(\psp)|$ and, in the binned approach, $F_{i}$.  (In recent LHCb analyses, though, $F_{i}$, and associated efficiency effects, have been obtained from simultaneous fits to \prt{B^{\pm} \to DK^{\pm}} and \prt{B^{\pm} \to D\pi^{\pm}} data~\cite{LHCb:2020yot, LHCb:2022nng}.)
\item \prt{D_2 \to g} is a \myCP-specific decay, either a \myCP-even decay such as \prt{D_2 \to K^+ K^-}, implying \prt{\ket{D_2} = \frac{1}{\sqrt{2}}\left( \ket{\Dz} + \ket{\Dzbar} \right)=: D_+}, or \myCP-odd such as \prt{D_2 \to \KS \pi^0}, implying 
\prt{\ket{D_2} = \frac{1}{\sqrt{2}}\left( \ket{\Dz} - \ket{\Dzbar} \right)=: D_-}, where the expressions for the superpositions of \Dz and \Dzbar are convention-dependent; we use a convention where $\myCP\ket{\Dz} = \ket{\Dzbar}$. The corresponding decay amplitudes are
\begin{align}
A(\psi \to D_1 D_{\pm}; D_1 \to f(\psp), D_{\pm} \to g) &\propto 
\mp \ADz^f(\psp) - \ADzbar^f(\psp),
\end{align}
and the decay rates:
\begin{align}
\label{eq:rateDCP}
\lefteqn{\Gamma(\psi \to D_1 D_{\pm}; D_1 \to f(\psp), D_2\to g)} &
\nonumber\\  &\propto 
|\ADz^f(\psp)|^2 + |\ADzbar^f(\psp)|^2 \pm 2|\ADz^f(\psp)| |\ADzbar^f(\psp)|\cos(\dd(\psp)).
\end{align}
These provide information on $\cos(\dd)$, or, in the binned approach, $c_i$.
\item Both \prt{D} mesons decay to the same self-conjugate signal mode e.g. $\KS \pi^+ \pi^-$:
\begin{align}
A(\psi \to D_1 D_2; D_1 \to f(\psp), D_2 \to f(\psq))
&\propto
\ADz(\psp)\ADzbar(\psq) - \ADzbar(\psp)\ADz(\psq),
\end{align}
with a decay rate
\begin{align}
\lefteqn
{\Gamma(\psi \to D_1 D_2; D_1 \to f(\psp), D_2 \to f(\psq)) } 
\;\;\;\;\;\;\;\;\;\;\;\;\;\;
&
\nonumber\\
& \propto 
|\ADz(\psp)|^2 |\ADzbar(\psq)|^2 + |\ADzbar(\psp)|^2 |\ADz(\psq)|^2
\nonumber\\ &\;\;\;\;\;
-2 |\ADz(\psp)||\ADzbar(\psq)| |\ADzbar(\psp)| |\ADz(\psq)|
\cos( \dd(\psp) - \dd(\psq) )
\\ &\propto
|\ADz(\psp)|^2 |\ADzbar(\psq)|^2 + |\ADzbar(\psp)|^2 |\ADz(\psq)|^2
\nonumber\\ 
\label{eq:rateDSvS}
&\;\;\;\;\;
-2 |\ADz(\psp)||\ADzbar(\psq)| |\ADzbar(\psp)| |\ADz(\psq)|
\nonumber\\ &\;\;\;\;\; \;\;\;\;
\times \left(
\cos( \dd(\psp))\cos(\dd(\psq) ) + \sin( \dd(\psp)) \sin(\dd(\psq) ) \right).
\end{align}
This category is therefore sensitive to both $\sin(\dd)$ and $\cos(\dd)$, in contrast to \myCP-tagged decays which are only sensitive to $\cos(\dd)$. In the binned approach, this translates into unique sensitivity to $s_i$.
\end{enumerate}
Measurements of $c_i$ and $s_i$ for \prt{D\to \KS \pi^+ \pi^-} have been made at \cleoc~\cite{CLEO:2010iul, CLEO:2009syb} and more recently \besIII~\cite{BESIII:2020hlg, BESIII:2020hpo}.

\section[Quasi Model Independent unbinned method to measure gamma]{Quasi Model Independent unbinned method to measure $\gamma$}
\label{sec:qmiMethodTheory}
\subsection{Basic idea}
This section will introduce the quasi model-independent (QMI) method that is the subject of this paper. For concreteness, we will consider the decay \prt{\Dz \to \KS \pi^+\pi^-}, although the method generalises to all self-conjugate charm decays, and with some modification also to non-self-conjugate ones.

The statistically most precise way of measuring $\gamma$ is the model-dependent method. We observe that the magnitude of the amplitude structure across the Dalitz plot of \prt{D^0 \to \KS \pi^+ \pi^-} is well known from flavour-specific \Dz\ and \Dzbar decays analysed by \babar and \belle~\cite{BaBar:2018cka}. The \Dz, \Dzbar datasets used in these fits are orders of magnitude larger than the \prt{B^{\pm} \to D K^{\pm}} datasets available for $\gamma$ measurements. 
The fact that the collaborations achieve a decent fit of their models to these datasets implies that we can trust the magnitude of existing amplitude models. However, those models violate unitarity and analyticity, which breaks the connection between magnitude and phase. 
Consequently, the phases of these models stand on less firm ground. 
Our approach is therefore to correct the model's phase in a model-independent way, or, more precisely, correct the \emph{phase difference} between the \Dz amplitude $\ADz(\psp)$ and \Dzbar amplitude $\ADzbar(\psp)$ at each phase space point \psp. These phase differences are all that matters and in fact it is all we have access to. Measurements of $c_i$ and $s_i$ by CLEO-c and BES~III suggest that the model's phase differences are at least approximately correct~\cite{BESIII:2020hlg,BESIII:2020hpo,CLEO:2010iul,CLEO:2009syb}. Our approach is therefore to obtain the model-independent phase difference by adding a correcting term \ddcorr to the model's phase difference \ddmodel, 
\begin{align}
\dd = \ddmodel + \ddcorr.
\end{align}
We parameterise \ddcorr in a generic way, as a power series in the Dalitz plot parameters. We assume \myCP conservation in charm decays, such that
\begin{align}
\dd(s_+, s_-) = -\dd(s_-, s_+),
\end{align}
which also implies
\begin{align}
\label{eq:ddcorrsymm}
\ddcorr(s_+, s_-) = - \ddcorr(s_-, s_+).
\end{align}
This symmetry reduces the number of parameters needed to parameterise \ddcorr.
We will see below that even if we depart from the assumption that the model's phase differences are approximately correct, such that \ddcorr becomes quite sizeable, our method still works.

The information that allows us to constrain \ddcorr comes, as for the binned methods, predominantly from the charm threshold, although \prt{B^{\pm} \to DK^{\pm}} decays also contribute.

\subsection[Constructing the correction to delta]{Constructing the correction to $\dd$}
\begin{figure}
    \begin{subfigure}{0.49\textwidth}
    \includegraphics[height=0.85\textwidth]{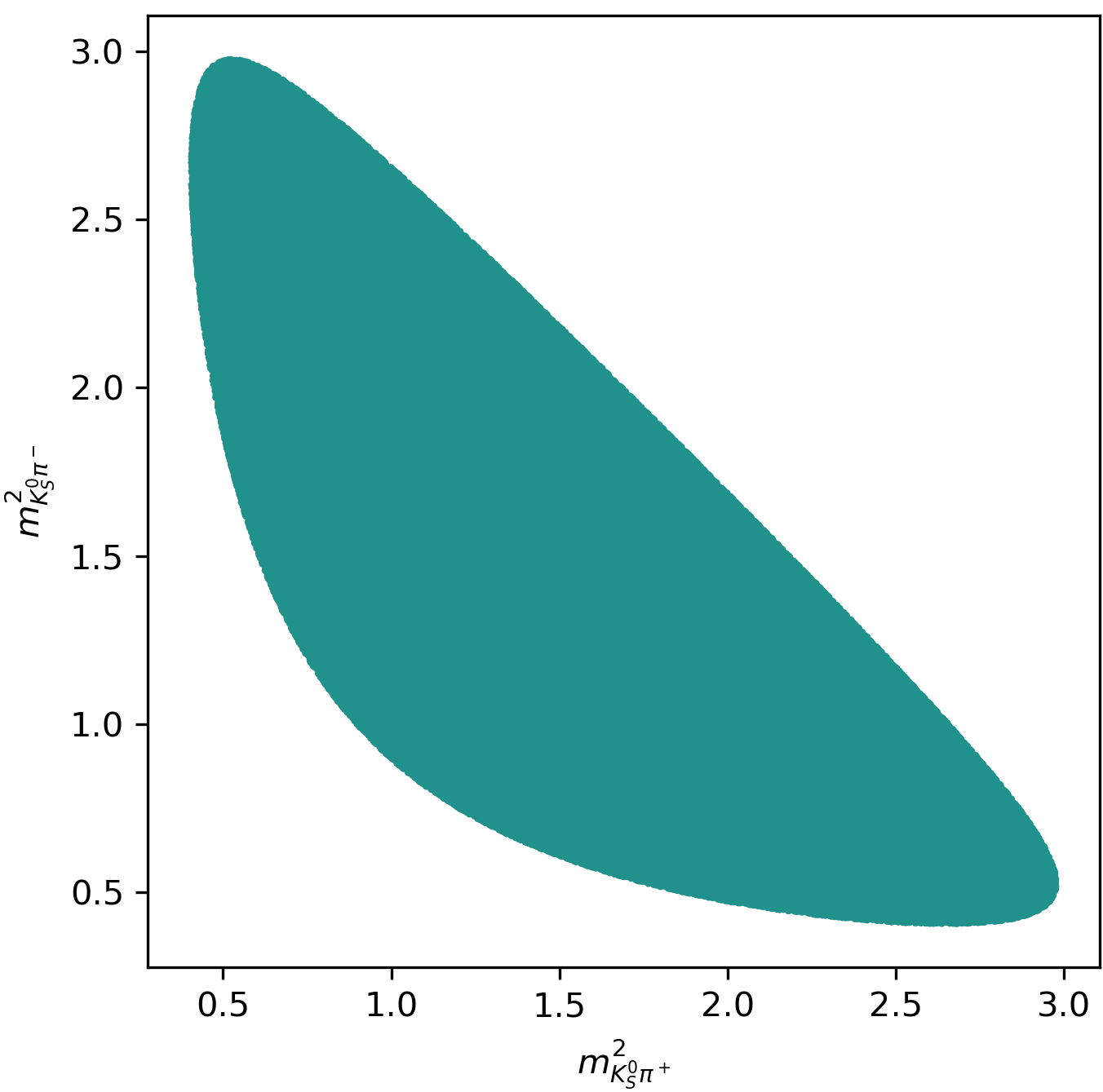}
    \caption{Dalitz Phasepace in terms of $s_+, s_-$}  
    \end{subfigure}
    \hfill
    \begin{subfigure}{0.49\textwidth}
    \includegraphics[height=0.85\textwidth]{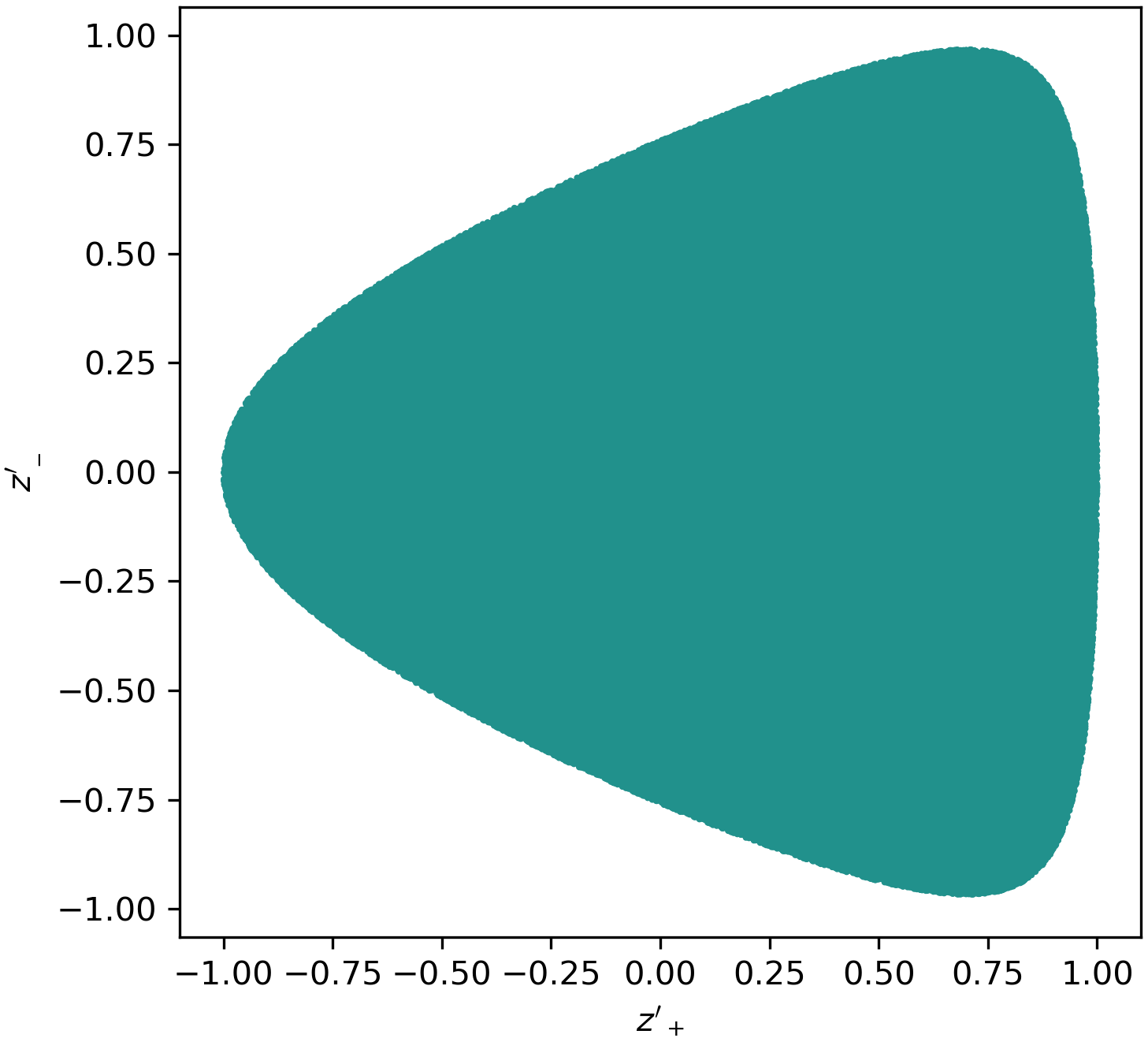}
    \caption{Rotated Dalitz Phasepace \label{fig:rotated}}  
    \end{subfigure}
    \mbox{}\vspace{3ex}\\
    \centering
    \begin{subfigure}{0.49\textwidth}
    \includegraphics[width=\textwidth]{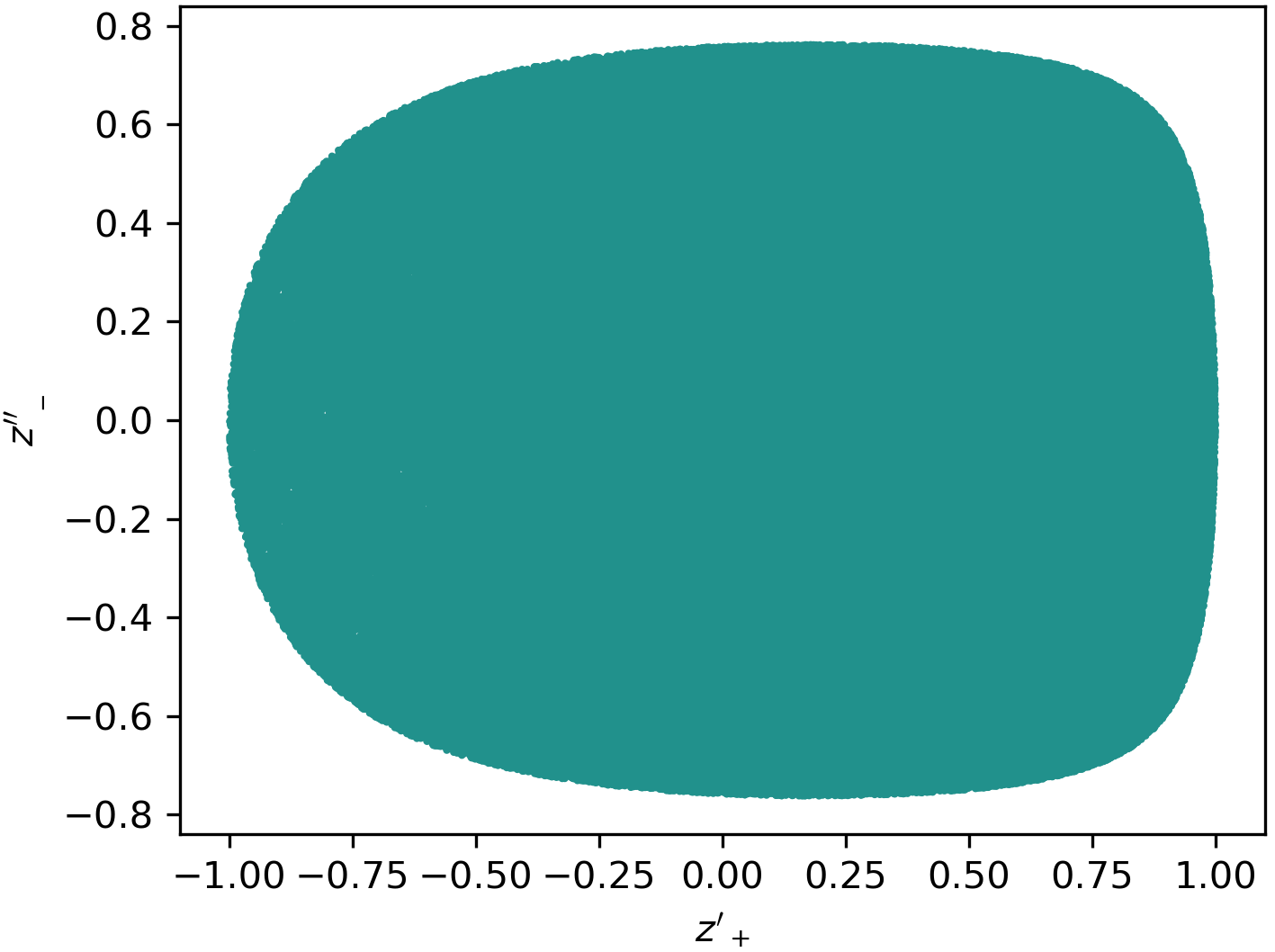}
    \caption{Stretched Rotated Dalitz Phasepace\label{fig:stretched_rotated}}  
    \end{subfigure}
    \caption{The Dalitz Phasespace, $(\mplus, \mminus)$, shaped to the rotated coordinates, $(z'_+, z'_-)$ and the stretched rotated coordinates, $(z'_+, z''_-)$.}
    \label{fig:setupPolyPhasespace}
\end{figure}
In order to implement the symmetry relation~\eqnref{eq:ddcorrsymm}, we define the variables 
\begin{align}
z_+ &\equiv s_+ + s_-, &
z_- &\equiv s_+ - s_-.
\end{align}
Now the symmetry condition from \eqnref{eq:ddcorrsymm} becomes
$\ddcorr(z_+, z_-) = - \ddcorr(z_+, -z_-)$ and can be implemented by only allowing terms  with odd powers of $z_-$ in the correcting polynomial.

We found that parameterising the phase in terms of Legendre polynomials works well. These are defined for values $x \in [-1, 1]$. We therefore scale $z_+, z_-$
\begin{align}
    z_+^{\prime} &= 
    \frac{2 z_+ - (z_+^{\mathrm{max}} + z_+^{\mathrm{min}})}{
    z_+^{\mathrm{max}}
    -z_+^{\mathrm{min}}
    },
    &
    z_-^{\prime} &= \frac{2 z_-
    - (z_-^{\mathrm{max}} + z_-^{\mathrm{min}})
    }{
    z_-^{\mathrm{max}}
    -z_-^{\mathrm{min}}
    },
\end{align}
where the superscripts max and min indicate the maximum and minimum values of the unprimed parameters. 
The kinematically allowed region of the \prt{\KS \pi^+ \pi^-} in terms of the normalised, rotated parameters $z_+^{\prime}, z_-^{\prime}$ is shown in \figref{fig:rotated}. 
The figure shows that within the square 
$z_+^{\prime} \in [-1,1], z_-^{\prime} \in [-1,1]$ 
where the Legendre polynomial is defined, there is a lot of space where there are no data, with the kinematically-allowed interval in $z_-^{\prime}$ varying a lot as a function of $z_+$. We therefore ``stretch'' the Dalitz plot by replacing $z_-^{\prime}$ with 
\begin{align}
    z_-^{\prime\prime} = \frac{2z_-^{\prime}}{z_+^{\prime} + 2}.
\end{align}
The Dalitz plot phase space for the different parametrisations is shown in \figref{fig:setupPolyPhasespace}. Alternatives such as a variation of the square Dalitz plot~\cite{Laura++} would have achieved a similar outcome.

We construct the correcting polynomial, $\ddcorr$, of order $N$ with free parameters $\mathbf{C}$
\begin{equation}
\ddcorr(z_+', z_-''| \mathbf{C}) = \sum_{i=0}^{i\leq N}\sum_{j=0}^{j \leq \frac{N-i}{2}} C_{i,2j+1}
 P_{i}(z'_+) P_{2j+1}(z''_-),
\end{equation}
where $P_n$ are, in our implementation, $n^{\mathrm{th}}$ order Legendre polynomials, although any function with $P_{2j+1}(z_-'') = -P_{2j+1}( -z_-'')$ would be equally valid.
The coefficients $C_{ij}$ are free parameters, determined together with $x_{\pm}, y_{\pm}$ in a simultaneous fit to $\psipp$ and \prt{B^{\pm} \to DK^{\pm}} data.

\section{Simulation Studies}
\label{sec:simulationStudies}
We test the algorithm in simulation studies where we generate charm threshold and \prt{B^{\pm} \to D K^{\pm}} data according to the equations given in \secref{sec:gammaFormalism} and \ref{sec:thresholdFormalism}, using the \prt{D^0\to \KS \pi^+ \pi^-} model in 
\ampgen~\cite{AmpGen}. We ignore backgrounds and efficiency effects in this study.

Using a modified version of AmpGen, we also generate data where we apply a change to the phase of the decay amplitude relative to the model assumed in the fit. 
We will then show how our new unbinned approach removes the bias this would inflict on the fitted value of $\gamma$ in a model-dependent approach, and how it does so with improved statistical uncertainty compared to the binned method.

\subsection{Simulated data}
\subsubsection{Sample sizes}
In our default settings, we simulate charm threshold data according to the event numbers given in BES~III's latest \prt{D\to \KS \pi^+ \pi^-} analysis \cite{BESIII:2020hpo} and \prt{B^{\pm} \to DK^{\pm}, D \to \KS\pi^+\pi^-} event yields reported in LHCb's latest measurement of $\gamma$ in this decay mode~\cite{LHCb:2020yot}. The BES~III signal yields for the different tags are shown in \tabref{tab:psippTags}.  For the purpose of this study, we treat \prt{\Dz \to K^- \pi^+} and \prt{\Dzbar \to K^+ \pi^-} as pure flavour tags and ignore the small contributions from \prt{\Dzbar \to K^- \pi^+} and \prt{\Dz \to K^+ \pi^-}.
\begin{table}
    \centering
    \begin{tabular}{l|r}
    Tag ($g$) & Events \\ \hline
    \myCP-even, e.g. \prt{D\to K^+ K^-}& $2546$  \\
     \myCP-odd, e.g. \prt{D \to \KS \pi^0} & $1725$ \\
    $\Dz$ flavour, e.g. \prt{D^0 \to K^- \pi^+} & $23457$ \\
    $\Dzbar$ flavour, e.g. \prt{\bar{D}^0 \to K^+ \pi^-} & $23457$ \\
    $D \to\kspipi$ (double tag) & $1833$
    \end{tabular}
    \caption{\prt{\psipp \to D_1 D_2 , D_1 \to \KS\pi^+\pi^-, D_2 \to g, } decays generated for our simulation studies. \prt{D_1}, \prt{D_2} represent superpositions of \Dz and \Dzbar, depending on the tag. \myCP even (odd) tags imply \myCP-odd (even) \prt{D \to \KS\pi^+\pi^-} decays and \Dz (\Dzbar) flavour tags imply \Dzbar (\Dz) decays to \prt{\KS \pi^+ \pi^-}. Sample sizes are taken from \cite{BESIII:2020hpo}. 
    \label{tab:psippTags}}
\end{table}
LHCb report a total of $12,533$ \prt{B^{\pm} \to DK^{\pm}, D \to \KS\pi^+\pi^-} decays. We split this evenly between \prt{B^- \to D K^-} and \prt{B^+ \to D K^+}, and generate on average $6267$ events for each. This implies that, as in LHCb's analysis,
our fits are entirely based on the distribution of events within each Dalitz plot, not \prt{B^+ \to DK^+} and \prt{B^- \to DK^-} event yields integrated across the whole Dalitz plot.
We also consider scenarios with larger datasets - by a factor of 100 for \prt{B^{\pm} \to DK^{\pm}, D \to \KS\pi^+\pi^-} and by a factor of 10 for charm threshold data.

\subsubsection{Amplitudes with modified phases}
We generate events based on the nominal amplitude model from \babar and \belle's joint analysis~\cite{BaBar:2018cka}, but with a modified phase difference
\begin{align}
    \ddmodel(\psp) \to \ddmodel(\psp) + f(\psp)
\end{align}
Note that the expressions for decay rates (\ref{eq:rateBm}, \ref{eq:rateBp}, \ref{eq:rateBmx} \ref{eq:rateBpx}, \ref{eq:rateDCP}, \ref{eq:rateDSvS}) only depend on the phase differences between \ADz and \ADzbar, never the absolute phases of \ADz or \ADzbar themselves. The function $f(\psp)$ modifies this phase difference. Our approach assures that the magnitudes $|\ADz|, |\ADzbar|$ remain unchanged. We generate events according to three scenarios:
\begin{enumerate}
    \item no phase modification, $f_0=0$,
    \item single Gaussian modification 
    \begin{equation}
    f_\mathrm{single}(\mplus, \mminus | A, \varepsilon, \mu_{+},\mu_{-}, 
    \sigma_{+}, \sigma_{-})
     = A~ \mathrm{erf}\left(\frac{\mplus - \mminus}{\varepsilon}\right) e^{-G\left(\mplus, \mminus \middle| \mu_\pm, \sigma_\pm\right)},
    \label{eqn:singlePeakBias}
     \end{equation}
     with
    \begin{equation}
    G\left(\mplus, \mminus\right) =
    \begin{cases}
    \frac{\left(\mplus - \mu_+\right)^2}{\sigma_+^2} + \frac{\left(\mminus - \mu_-\right)^2}{\sigma_-^2} & \mplus > \mminus, \\
    \frac{\left(\mminus - \mu_+\right)^2}{\sigma_+^2} + \frac{\left(\mplus - \mu_-\right)^2}{\sigma_-^2} & \mplus < \mminus, 
    \end{cases}
    \end{equation}
    \item double Gaussian modification
    \begin{equation}
    \label{eq:doubleBias}
    \begin{split}
    f_\mathrm{double}(\mplus, \mminus) &= f_\mathrm{single}(\mplus, \mminus | A_1, \varepsilon_1, \mu_{+1},\mu_{-1}, \sigma_{+1}, \sigma_{-1}) \\ &+  f_\mathrm{single}(\mplus, \mminus | A_2, \varepsilon_2, \mu_{+2},\mu_{-2}, \sigma_{+2}, \sigma_{-2}).\\  
    \end{split}
    \end{equation}
\end{enumerate}
\begin{table}
\centering
\begin{tabular}{c|rrrrrr}
                             & $A$ & $\varepsilon$ 
                             & $\mu_{+}$ & $\sigma_{+}$ &$\mu_{-}$ & $\sigma_{-}$ \\ \hline
     $f_{\mathrm{single}}  $           &   1  & 0.1 & 2.0 & 0.75 & 0.90 & 0.25\\
     $f_{\mathrm{double}}$ subscript 1 &   1  & 0.1 & 1.0 & 0.25 & 1.25 & 1.0\\
     $f_{\mathrm{double}}$ subscript 2 & $-1$ & 0.1 & 2.5 & 0.25 & 1.25 & 1.0 
\end{tabular} 
\caption{Parameters used for scenarios two ($f_{\mathrm{single}}$) and three ($f_{\mathrm{double}}$), where for $f_{\mathrm{double}}$ the top row refers to the parameters with subscript $1$ in \eqnref{eq:doubleBias}, and the bottom row refers those with subscript $2$. The parameters $\varepsilon$, $\mu$ and $\sigma$ are given in units of $\left(\mathrm{GeV}/c^2\right)^2$.\label{tab:biasParameters}}
\end{table}
The purpose of the error function, $\mathrm{erf(x)} \equiv \frac{2}{\sqrt{\pi}}\int_0^x e^{-t^2}\,dt$, in the definition of $f_{\mathrm{single}}$ is to implement the condition $f(s_+, s_-) = -f(s_-, s_+)$ while providing a smooth transition across the line $s_+ = s_-$.
\begin{figure}
    \centering
    \begin{subfigure}[t]{0.49\textwidth}
    \includegraphics[width=\textwidth]{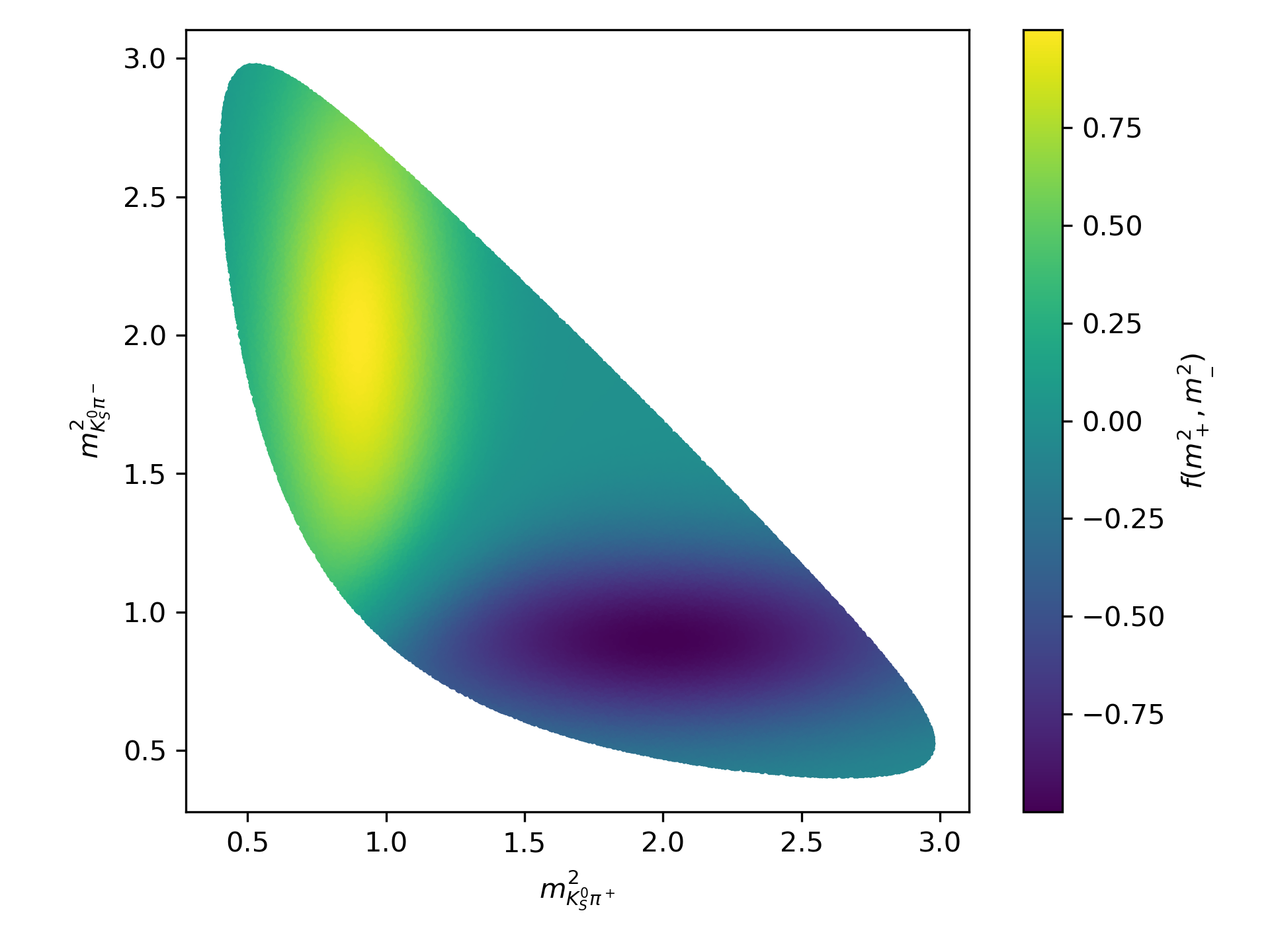}
    \caption{Single peak bias
    \label{fig:singlePeakBias}}
    \end{subfigure}
    \hfill
    \begin{subfigure}[t]{0.49\textwidth}
    \includegraphics[width=\textwidth]{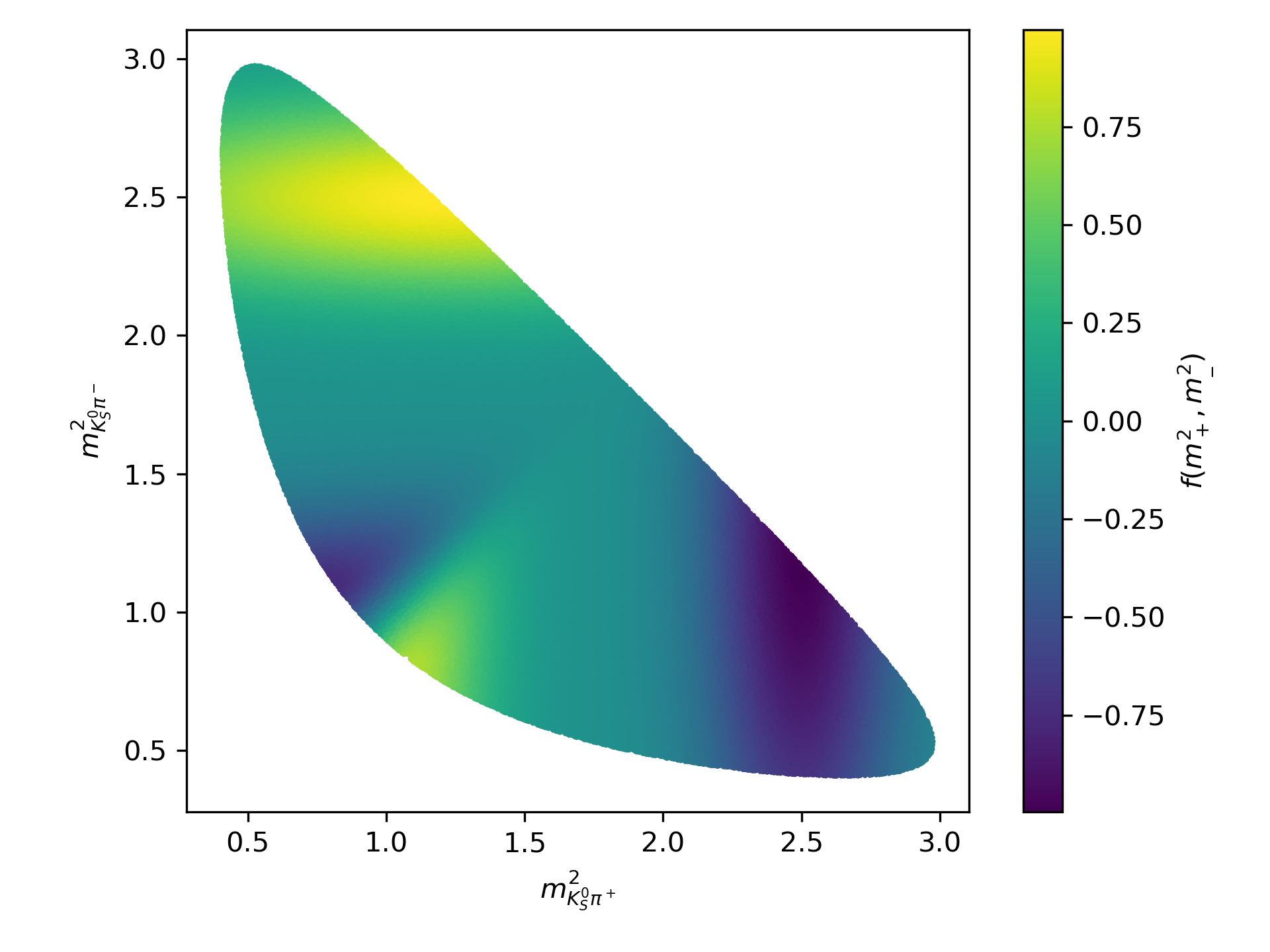}
    \caption{Double peak bias
    \label{fig:doublePeakBias}}
    \end{subfigure}
    \caption{The figures show the difference between the phase difference $\dd(s_+, s_-)$ of the nominal amplitude model and the model with which the data are actually generated, for the two biased scenarios considered.
    \label{fig:biases}}
\end{figure}
The parameter values used for scenarios two and three are given in \tabref{tab:biasParameters}. The phase modifications they induce are shown in \figref{fig:biases}, which shows $f_{\mathrm{single}}$ and $f_{\mathrm{double}}$, in radians. 
It can be seen that scenario two targets the region of the \prt{K^*} resonance, while scenario three has large phase modifications especially in the region of the \prt{\rho \KS -K^*\pi} interference. 
It is worth noting that the phase change relative to the nominal model we consider here is up to $\pm 1$ radian, which is not a small shift.

\subsubsection{Other input parameters}
All samples are generated with the 2022 HFLAV~\cite{HFLAV} averages $r_B=0.093, \delta_B = 119.5^{\circ}, \gamma= 69.5^{\circ}$, which corresponds to
\begin{align}
        x_+ &= -0.09186, & y_+  &= -0.01455; &
        x_- &= 0.05978,  & y_- &= 0.07124.
\end{align}
\subsection{Fit results}
\subsubsection{Order by order fits to individual samples}
We perform fits with the model-dependent (MD) method and with the quasi model-independent (QMI) method with phase-correction polynomials of order $N =1, 2, \ldots, 9$, for one sample of each of the phase-modification scenarios. Beyond $N=9$, fits converge very slowly due to the large number of parameters.
The results are shown in tables~\ref{tab:orderToOrderNoBias}, \ref{tab:orderToOrderSinglePeak}, and \ref{tab:orderToOrderDoublePeak}, in the format (fit result) $-$ (input value) $\pm$ (uncertainty), in units of $10^{-2}$;  the uncertainty is that estimated by the MINUIT2~\cite{James:1975dr}-based AmpGen~\cite{AmpGen} fitter (validated below, in \secref{sec:validation}). 
We see in \tabref{tab:orderToOrderNoBias}, where there is no phase modification relative to the nominal amplitude model, that the model-dependent and the quasi-model-independent fit both reproduce the input values, within uncertainties. 
For correction polynomials of order $N>3$, the fit results differ slightly between the methods, but far less than the statistical uncertainty. 
The validation studies below show that this does not lead to a systematic bias. Tables~\ref{tab:orderToOrderSinglePeak} and~\ref{tab:orderToOrderDoublePeak} show that the phase modifications induce a significant bias in $x_{\pm}, y_{\pm}$ in the model-dependent method, and how the QMI method recovers from it. 
The changes in the fit results for $x_{\pm}, y_{\pm}$ between the model-dependent and the higher order QMI fits correspond to changes in the estimated values of $r_B$, $\delta_B$, and $\gamma$ of $0.006$, $11^{\circ}$, and $7^{\circ}$ for the double-Gaussian bias (\tabref{tab:orderToOrderDoublePeak}), and of $0.021$, $23^{\circ}$, and $0.3^{\circ}$ for the single-Gaussian bias (\tabref{tab:orderToOrderSinglePeak}).
An interesting feature is that the uncertainties on $x_{\pm}$ and $y_{\pm}$ do not appear to be affected significantly by the additional fit parameters. This conclusion is confirmed in the validation studies shown below.
\begin{table}
    \centering
    \begin{tabular}{c|c|c|c|c|c|c|c}
 Order & $\Delta x_+ \cdot 100$ & $\Delta y_+\cdot 100$ & $\Delta x_-\cdot 100$ & $\Delta y_-\cdot 100$  \\ 
\hline
MD&$-0.9\pm0.8$&$-1.1\pm 1.1$&$-1.5\pm0.9$&$+1.0\pm0.9$\\
1& $-0.9\pm0.8$&$-1.0\pm 1.1$&$-1.5\pm0.9$&$+0.9\pm0.9$\\
2& $-0.9\pm0.8$&$-1.0\pm 1.1$&$-1.5\pm0.9$&$+1.0\pm0.9$\\
3& $-0.9\pm0.8$&$-1.2\pm 1.1$&$-1.5\pm0.9$&$+1.1\pm0.9$\\
4& $-0.8\pm0.8$&$-1.1\pm 1.1$&$-1.6\pm0.9$&$+1.2\pm0.9$\\
5& $-0.9\pm0.8$&$-1.1\pm 1.1$&$-1.6\pm0.9$&$+1.2\pm0.9$\\
6& $-0.9\pm0.8$&$-1.1\pm 1.2$&$-1.5\pm0.9$&$+1.1\pm0.9$\\
7& $-0.9\pm0.8$&$-1.1\pm 1.2$&$-1.5\pm0.9$&$+1.1\pm0.9$\\
8& $-0.8\pm0.8$&$-1.3\pm 1.2$&$-1.6\pm0.9$&$+1.3\pm0.9$\\
9& $-0.8\pm0.8$&$-1.4\pm 1.2$&$-1.6\pm0.9$&$+1.3\pm0.9$
    \end{tabular}
    \caption{Order to order fit for the unbiased $\dd$ sample.}  
    \label{tab:orderToOrderNoBias}
\end{table}
\begin{table}
    \centering
\begin{tabular}{c|c|c|c|c}
 Order & $\Delta x_+ \cdot 100$ & $\Delta y_+\cdot 100$ & $\Delta x_-\cdot 100$ & $\Delta y_-\cdot 100$  \\ \hline
 MD& $-1.8\pm0.8$&$+4.0\pm 1.1$ &$+1.0\pm0.9$&$-3.7\pm0.9$\\
1  & $-0.6\pm0.8$&$+1.0\pm 1.0$ &$+1.4\pm0.9$&$-0.8\pm0.9$\\
2  & $-0.7\pm0.8$&$+0.9\pm 1.0$ &$+1.4\pm0.9$&$-0.7\pm0.9$\\
3  & $-0.3\pm0.8$&$+0.3\pm 1.1$ &$+1.1\pm0.9$&$+0.5\pm1.0$\\
4  & $-0.3\pm0.8$&$+0.2\pm 1.0$ &$+1.1\pm0.9$&$+0.7\pm1.0$\\
5  & $-0.2\pm0.8$&$+0.1\pm 1.1$ &$+1.1\pm0.9$&$+0.9\pm1.0$\\
6  & $-0.2\pm0.8$&$-0.1\pm 1.0$ &$+1.0\pm0.9$&$+0.9\pm1.0$\\
7  & $-0.2\pm0.8$&$-0.2\pm 1.0$ &$+1.0\pm0.9$&$+1.0\pm1.0$\\
8  & $-0.2\pm0.8$&$-0.3\pm 1.0$ &$+1.1\pm0.9$&$+0.9\pm1.0$\\
9  & $-0.2\pm0.8$&$-0.2\pm 1.0$ &$+1.1\pm0.9$&$+0.9\pm1.0$
    \end{tabular}
        \caption{Order to order fit for $\dd + f_\mathrm{single}(\mplus, \mminus)$ sample.}
    \label{tab:orderToOrderSinglePeak}
\end{table}
\begin{table}
    \centering
\begin{tabular}{c|c|c|c|c}
 Order & $\Delta x_+ \cdot 100$ & $\Delta y_+\cdot 100$ & $\Delta x_-\cdot 100$ & $\Delta y_-\cdot 100$  \\ \hline
 MD&$+1.3\pm0.8$&$+1.2\pm 1.1$&$-1.0\pm 1.3$&$-3.3\pm 1.3$\\
1  &$+1.1\pm0.8$&$+0.5\pm 1.0$&$-1.3\pm0.8$&$-0.6\pm 1.0$\\
2  &$+0.5\pm0.9$&$+0.1\pm 1.0$&$-1.0\pm0.8$&$+0.4\pm 1.0$\\
3  &$+0.6\pm0.8$&$\;\;0.0\pm 1.0$&$-1.2\pm0.8$&$+0.4\pm 1.0$\\
4  &$+0.3\pm0.8$&$+0.4\pm 1.0$&$-0.8\pm0.8$&$+0.3\pm 1.0$\\
5  &$+0.4\pm0.8$&$+0.5\pm 1.0$&$-0.7\pm0.8$&$+0.3\pm 1.0$\\
6  &$+0.3\pm0.8$&$+0.7\pm 1.0$&$-0.8\pm0.8$&$+0.4\pm 1.0$\\
7  &$+0.3\pm0.8$&$+0.5\pm 1.0$&$-0.9\pm0.8$&$+0.7\pm 1.0$\\
8  &$+0.3\pm0.8$&$+0.5\pm 1.0$&$-0.9\pm0.8$&$+0.7\pm 1.0$\\
9  &$+0.3\pm0.8$&$+0.5\pm 1.0$&$-0.7\pm0.8$&$+0.7\pm 1.0$
    \end{tabular}
        \caption{Order to order fit for $\dd + f_\mathrm{double}(\mplus, \mminus)$ sample.}
        \label{tab:orderToOrderDoublePeak}
\end{table}

\subsubsection{Fits to 100 pseudoexperiments}
\label{sec:validation}
\begin{table}
\centering
\begin{tabular}{cc|cccc}
phase
     & Method
     & $(\Delta x_+ \pm \sigma_{x_+})$ 
     & $(\Delta y_+ \pm \sigma_{y_+})$
     & $(\Delta x_- \pm \sigma_{x_-})$
     & $(\Delta y_- \pm \sigma_{y_-})$ \\
mod.
     & 
     & $\times 100$
     & $\times 100$
     & $\times 100$
     & $\times 100$
     \\ \hline
\multirow{2}{*}{$f_0=0$} &
QMI  & $-0.1\pm 0.6$ & $+0.1 \pm 1.1$ & $-0.1\pm 1.0$ & $\;\;\,0.0\pm 1.0$ \\
& MD  & $-0.1\pm 0.6$ & $+0.1 \pm 1.1$ & $-0.1\pm 0.9$ & $-0.1\pm 1.0$ 
\\ \hline
 \multirow{2}{*}{$f_{\mathrm{single}}$} &
 QMI & $+0.1 \pm 0.7$&$+0.1 \pm 1.1$&$+0.4 \pm 0.9$&$-0.1 \pm 1.1$ \\
& MD &  $+0.9 \pm 0.7$&$+3.6 \pm 1.1$&$+0.3 \pm 0.9$&$-3.7 \pm 1.2$ 
\\ \hline
 \multirow{2}{*}{$f_{\mathrm{double}}$} &
 QMI & $+0.1 \pm 0.7$&$-0.0 \pm 1.0$&$+0.1 \pm 0.9$&$+0.2 \pm 1.0$ \\
& MD &  $+0.5 \pm 0.7$&$+1.8 \pm 1.1$&$+0.1 \pm 1.0$&$-1.6 \pm 1.0$ \\
\end{tabular}
\caption{Residuals from 100 fits to samples without any phase modification, with the quasi-model-independent (QMI) and the model-dependent (MD) method. The QMI fit uses a 6th order phase-correction polynomial. Results are shown in the format (mean result) $-$ (input value) $\pm$ (standard deviation), in units of $10^{-2}$. The standard deviation $\sigma$ is that of the distribution of fit results; the uncertainty on the mean is $\sigma/\sqrt{100}$. The uncertainty on the standard deviation, $\sigma/\sqrt{200}$, varies between $0.04 \times 10^{-2}$ and $0.08\times 10^{-2}$.
\label{tab:residuals}}
\end{table}
\begin{table}
\centering
\begin{tabular}{cc|cccc}
    phase-mod
     & Method
     & $\frac{\Delta x_+}{\sigma_{x_+}}$ 
     & $\frac{\Delta y_+}{\sigma_{y_+}}$
     & $\frac{\Delta x_-}{\sigma_{x_-}}$
     & $\frac{\Delta y_-}{\sigma_{y_-}}$
     \\ \hline
\multirow{2}{*}{$f_0=0$} &
QMI  & $-0.12\pm 0.82$ & $+0.07 \pm 1.01$ & $-0.12\pm 1.11$ & $-0.02\pm 1.06$ \\
& MD  & $-0.08\pm 0.82$ & $+0.13 \pm 1.01$ & $-0.07\pm 1.06$ & $-0.12\pm 1.01$ 
\\ \hline
 \multirow{2}{*}{$f_{\mathrm{single}}$} &
QMI   & $+0.22 \pm 0.89 $ & $ -0.07 \pm 0.93$ & $  +0.16 \pm 1.08 $ & $+0.07\pm 1.28 $ \\
& MD  & $+1.10 \pm 0.85$ & $ +3.42 \pm 1.03 $ & $  +0.36 \pm 1.03 $ & $-3.85 \pm 1.21$ 
\\ \hline
 \multirow{2}{*}{$f_{\mathrm{double}}$} &
QMI   & $ +0.17\pm 0.90 $ & $ +0.07 \pm 0.94$ & $+0.02 \pm 0.99$ & $ +0.13\pm 1.01$ \\ 
& MD  & $ +2.04\pm 0.87 $ & $ +1.07 \pm 0.95$ & $-0.93\pm 1.16$ & $-1.81 \pm 1.24$ \\
\end{tabular}
\caption{Pull results from 100 fits with the quasi-model-independent (QMI) and the model-dependent (MD) method, for each of the three phase-modification scenarios. The QMI fit uses a 6$^{\mathrm {th}}$ order correction polynomial. Results are shown in the format (mean pull) $\pm$ (standard deviation). The standard deviation is that of the pull distribution (rather than the uncertainty on the mean). The uncertainty on the mean is $0.1$, that on the standard deviation is $0.07$. The substantial (and expected) biases observed for the model-dependent method for the fits with phase-modification disappear with the QMI method.  
\label{tab:pulls}}
\end{table}

We generate 100 pseudoexperiments and fit them with the model-dependent method, the binned method, and the QMI method. The QMI method uses a 6\te order correction polynomial. 
\Tabref{tab:residuals} shows the mean and standard deviation of the distribution of residuals (i.e. fit result minus truth value) for these fits. 
\Tabref{tab:pulls} shows the corresponding value for the pull, which is the residual divided by the uncertainty reported by the fit. The results show that, in the absence of any phase modification, the model-dependent and the QMI method both yield unbiased results with essentially the same uncertainty. 
The pulls for the QMI method, and for the unbiased data also those for the model-dependent method, show generally good agreement with the expected mean of zero and standard deviation of one. 
For both methods, the uncertainty the fitter reports on $x_+$ seems to be slightly over-estimated. There appears to be a slight under-estimation of the $y_-$ uncertainty in the $f_{\mathrm{single}}$ configuration. The results with the two phase modification scenarios confirm that the phase modifications induce significant biases in the fit results of the model-dependent method, while the QMI results remain unbiased.

\begin{table}
    \centering
\begin{tabular}{c|cccc|ccc}
     & $\sigma_{x_+}$ & $\sigma_{y_+}$ & $\sigma_{x_-}$ & $\sigma_{y_-}$ 
     & $ \sigma_{r_B} $ & $\sigma_{\delta_B} $ & $\sigma_{\gamma}$ \\
     & $\times 10^2$  & $\times 10^2$  & $\times 10^2$   & $\times 10^2$ 
     & $\times 10^2$  &\\
\hline
binned fit (fixed $c_i, s_i$) & 
$0.886$&$1.482$&$1.189$&$1.328$& $0.879$ & $5.33^{\circ}$ 
& $5.09^{\circ}$ 
\\ 
unbinned QMI &
$0.780$&$1.091$&$0.877$&$0.945$&$0.664$ &$4.24^{\circ}$ 
&$4.21^{\circ}$ 
\\
unbinned MD &
$0.784$&$1.081$&$0.878$&$0.939$&$0.660$
&$4.19^{\circ}$ 
&$4.23^{\circ}$ 
\\
\end{tabular}
    \caption{Comparing the QMI method with our implementation of the model-dependent method and the binned method with ``optimal'' binning (defined in~\cite{CLEO:2010iul}), 
    for the case with no phase modification. The uncertainties given are the average of those reported by the fitter for $100$ fits. 
    For the binned fit, we fix $c_i$ and $s_i$. 
    In contrast to the QMI results, the uncertainties from the binned fit therefore do not include the effect of the finite sample size at the charm threshold, which leads to an additional uncertainty on $\gamma$ of $1.2^{\circ}$~\cite{BESIII:2020hlg}. 
    \label{tab:compareWithBinned}}
\end{table}
\Tabref{tab:compareWithBinned} compares the uncertainties obtained with our new method to those from 
the model-dependent and the model-independent binned method. 
Studies in~\cite{Poluektov:2017zxp} show that the unbinned model-independent method introduced there, 
which is based on projecting the two-dimensional Dalitz plot onto one dimension, 
results in a statistical uncertainty on $\gamma$ between that of the model-dependent and the binned method.
The authors of~\cite{Backus:2022xhi} report for their Kolmogorov-Smirnov-inspired unbinned method, for similar simulated event numbers as used in our studies, a statistical uncertainty on $\gamma$ of~$\sim 5^{\circ}$. 
However, because of the different values assumed for $\gamma$ and $\delta_B$ 
and differences in the implementation of the amplitude model, comparing the results from~\cite{Backus:2022xhi} with those in \tabref{tab:compareWithBinned} is not straightforward.

In our implementation of the binned method, we base the binning on the same amplitude model that we use to generate the simulated data, which should result in a slightly optimistic performance of the binned method. 
We test all binning schemes defined  in~\cite{CLEO:2010iul} and find that the ``optimal'' binning scheme leads to the best results.  
In our binned fit, we fix $c_i$ and $s_i$ to their true value (according to our model), so that the uncertainty on $\gamma$ for the binned method does not include the contribution from the uncertainty on 
$c_i$ and $s_i$. 
The sensitivity studies reported in~\cite{BESIII:2020hpo} show that, for the ``optimal'' binning scheme, taking into account the measurement uncertainties on $c_i$ and $s_i$ leads to an additional uncertainty on $\gamma$ of $1.2^{\circ}$. This results in a total uncertainty on $\gamma$ of $5.1^{\circ} \oplus 1.2^{\circ} = 5.2^{\circ}$, which is, on the same simulated signal data, improved to $4.2^{\circ}$ by the new method introduced, here.

\subsubsection{Alternative sample sizes}
For the studies above, we used sample sizes corresponding to those reported in recent publications by \besIII~\cite{BESIII:2020hpo, BESIII:2020hlg} and LHCb~\cite{LHCb:2020yot}. Here we consider possible future datasets that are considerably larger, $10\times$ as large for \besIII and $100\times$ as large for LHCb.
\begin{table}
    \centering
         \begin{tabular}{r *{5}{|ll}}
    LHCb & 
    \multicolumn{2}{c|}{$ \sigma_{x_+} \cdot 10^2 $} &
    \multicolumn{2}{c|}{$ \sigma_{y_+} \cdot 10^2 $} &
    \multicolumn{2}{c|}{$ \sigma_{x_-} \cdot 10^2 $} &
    \multicolumn{2}{c|}{$ \sigma_{y_-} \cdot 10^2 $} &
    \multicolumn{2}{c}{$ \sigma_{\gamma}$ (${}^\circ $)}
    \\ \cline{2-11}
    Lumi & MD & bin & MD & bin & MD & bin & MD & bin  & MD & bin 
    \\ \hline
    $\times 1$ 
&$0.780$ & $0.886$
&$1.081$ & $1.482$
&$0.878$ & $1.189$
&$0.939$ & $1.328$
&$4.23$ 
&$5.09$ 
    \\
    $\times 100$
&$0.078$ & $0.089$ 
&$0.108$ & $0.149$
&$0.088$ & $0.118$
&$0.093$ & $0.134$
&$0.42$ 
&$0.52$ 
     \end{tabular}
    \caption{Uncertainties on fit parameters for the model-dependent method and the binned method with fixed $c_i, s_i$ for $1\times$ and $100\times$ the dataset analysed by LHCb in~\cite{LHCb:2020yot}. The uncertainties are the average of those reported by the fitter for fits to 100 pseudoexperiments, generated without backgrounds or detector effects. The statistical uncertainty on the mean of $\sigma_{\gamma}$ ranges from $1\%$ to $3\%$ of its value.
    \label{tab:lumiscenariosMDandBinned}}
\end{table}
\begin{table}
    \centering
\begin{tabular}{rr *{5}{|c}}
    \multicolumn{2}{c|}{Lumi scenario:}  \\
        LHCb & \besIII &
    \multicolumn{1}{c|}{$ \sigma_{x_+} \cdot 10^2 $} &
    \multicolumn{1}{c|}{$ \sigma_{y_+} \cdot 10^2 $} &
    \multicolumn{1}{c|}{$ \sigma_{x_-} \cdot 10^2 $} &
    \multicolumn{1}{c|}{$ \sigma_{y_-} \cdot 10^2 $} &
    \multicolumn{1}{c}{$ \sigma_{\gamma}$ (${}^\circ $)}
    \\ \hline
    $\times 1$ & $\times 1$
&$0.780$
&$1.091$
&$0.877$
&$0.945$
&$4.21$ 
    \\
    $\times 1$ & $\times 10$
&$0.773$
&$1.062$
&$0.866$
&$0.924$
&$4.18$ 
    \\
    $\times 100$ & $\times 1$
&$0.079$
&$0.122$
&$0.090$
&$0.104$
&$0.45$ 
    \\
    $\times 100$ & $\times 10$
&$0.078$
&$0.115$
&$0.089$
&$0.099$
&$0.43$ 
    \\
     \end{tabular}
    \caption{Uncertainties on fit parameters for the QMI method, for scenarios with $1\times$ and $100\times$ the dataset analysed by LHCb in~\cite{LHCb:2020yot}, and $1\times$ and $10\times$ the dataset analysed by \besIII in~\cite{BESIII:2020hpo,BESIII:2020hlg}. 
    The uncertainties are the average uncertainty reported by the fitter for $\sim 100$ simulated datasets, generated without backgrounds or detector effects. The statistical uncertainty on the mean of $\sigma_{\gamma}$ is $\sim 1\%$ of its value.    \label{tab:lumiscenariosQMI}}
\end{table}
The results for the QMI method are presented in \tabref{tab:lumiscenariosQMI}. The results for the model-dependent method and the binned method with fixed $c_i, s_i$ are given in \tabref{tab:lumiscenariosMDandBinned}. 
The results for the binned method, with fixed $c_i, s_i$, represent the best possible uncertainty on $\gamma$ that can be reached with this method in the limit of infinitely large threshold data sets, given the 8-bin-pair binning scheme and other parameters used in this study. It is possible that a finer binning would improve the uncertainty for the larger data set. 

The uncertainties for all methods studied scale to a good approximation with $1/\sqrt{N_B}$, where $N_B$ is the number of \prt{B^{\pm} \to DK^{\pm}, D \to \KS \pi^+ \pi^-} events.
That this is so for the QMI method is not a priori obvious, given that it depends also on \prt{\psi(3770) \to D\overline{D}} (i.e. CLEO-c or \besIII) data. This suggests that the QMI method is very efficient in extracting information on $\dd(s_+, s_-)$ not only from threshold data, but also from  \prt{B^{\pm} \to DK^{\pm}, D \to \KS \pi^+ \pi^-}. 

For the $1\times$LHCb scenario, the lack of significant improvement in the uncertainty on $\gamma$ for the $10\times$ larger \besIII sample is consistent with the earlier observation that the QMI method with the $1\times$LHCb and $1\times$\besIII scenario already achieves effectively the same statistical precision as the model-dependent method (i.e. the best possible for the \prt{B^{\pm} \to DK^{\pm}, D \to \KS \pi^+ \pi^-} dataset). 
For the $100 \times$LHCb dataset, the larger \besIII sample improves the precision slightly from $0.45^{\circ}$ to $0.43^{\circ}$, compared to the benchmark of $0.42^{\circ}$ set by the model-dependent method.

Overall, our results indicate that, with the QMI method introduced here, the statistical precision on $\gamma$ in \prt{B^{\pm} \to D K^{\pm}, D\to \KS \pi^+ \pi^-} remains close to the optimum defined by the model-dependent method not only with current datasets, but also much larger \prt{B^{\pm} \to D K^{\pm}} datasets that might become available in the future.

\section{Conclusion}
\label{sec:conclusion}
We have introduced a novel unbinned quasi model-idependent (QMI) method for the measurement of $\gamma$ in \prt{B^{\pm} \to D K^{\pm}} decays, with input from quantum-correlated charm threshold data.
The method uses a polynomial to correct the phase of the \prt{D} meson's decay amplitude model in an unbinned, model-independent way.

We studied the performance of the method using simulated \prt{B^{\pm} \to D K^{\pm}, D\to \KS \pi^+ \pi^-} and $\psi(3770) \to D\overline{D}$ signal events. The method produces unbiased results for cases where discrepancies between the assumed amplitude model and the true model produce large biases in a model-dependent measurement. 
For realistic current and plausible future sample sizes, the method achieves a statistical precision on $\gamma$ that is effectively the same as the optimum defined by the model-dependent method, without suffering from the systematic uncertainty associated with the amplitude model. The statistical uncertainty is significantly better than that of the binned model-independent method currently in use.

We expect that the QMI method will also be beneficial in the study of charm mixing and the study of phases of decay amplitudes across the Dalitz plot.

\acknowledgments
J. Lane and E. Gersabeck have received support by the \href{https://royalsociety.org/}{Royal Society} (UK), J. Rademacker from \href{https://www.ukri.org/councils/stfc}{STFC} (UK); we express our gratitude. 
We thank Dr Tim Evans for the development and support of the Amplitude Analysis framework \ampgen~\cite{AmpGen}, upon which our software is built. We thank Prof Marco Gersabeck, Dr Martha Hilton (now at the \href{https://www.iop.org/}{Institute of Physics} (UK)), Dr Mark Williams (now at the University of Edinburgh), Dr Florian Reiss, and Dr David Friday from the University of Manchester; Prof Tim Gershon from the University of Warwick; and Jozie Cottee Meldrum from the University of Bristol  for their insightful feedback. 
We are indebted to the participants of the
\href{https://inspirehep.net/conferences/2096641}{Charming Clues for Existence workshop} (2022) at the Munich Institute for Astro and Particle Physics (supported by the German Research Foundation, 
DFG, project \href{https://gepris.dfg.de/gepris/projekt/390783311?language=en}{EXC-2094 – 390783311}) for helpful comments and discussions.
\bibliography{refs}

\end{document}